\documentclass{article}

\usepackage[final]{neurips_2024}

\usepackage[utf8]{inputenc} 
\usepackage[T1]{fontenc}    
\usepackage{hyperref}       
\usepackage{url}            
\usepackage{booktabs}       
\usepackage{amsfonts}       
\usepackage{nicefrac}       
\usepackage{microtype}      
\usepackage{xcolor}         
\usepackage{amsmath}

\setcitestyle{numbers,square,citesep={,}}
\usepackage{graphicx}
\usepackage{float}

\title{F5R-TTS: Improving Flow-Matching based Text-to-Speech with Group Relative Policy Optimization}

\author{%
  Xiaohui Sun, Ruitong Xiao, Jianye Mo, Bowen Wu, Qun Yu, Baoxun Wang \\
  Platform and Content Group, Tencent\\
  \texttt{\{bryanxsun, rayrtxiao, kevinmo, jasonbwwu, sparkyu, asulewang\}@tencent.com} \\
}

\begin{document}

\maketitle

\begin{abstract}
  
  We present F5R-TTS, a novel text-to-speech (TTS) system that integrates Group Relative Policy Optimization (GRPO) into a flow-matching based architecture. By reformulating the deterministic outputs of flow-matching TTS into probabilistic Gaussian distributions, our approach enables seamless integration of reinforcement learning algorithms. During pretraining, we train a probabilistically reformulated flow-matching based model which is derived from F5-TTS with an open-source dataset. In the subsequent reinforcement learning (RL) phase, we employ a GRPO-driven enhancement stage that leverages dual reward metrics: word error rate (WER) computed via automatic speech recognition and speaker similarity (SIM) assessed by verification models. 
  Experimental results on zero-shot voice cloning demonstrate that F5R-TTS achieves significant improvements in both speech intelligibility (a 29.5\% relative reduction in WER) and speaker similarity (a 4.6\% relative increase in SIM score) compared to conventional flow-matching based TTS systems.
  Audio samples are available at \url{https://frontierlabs.github.io/F5R}.

\end{abstract}

\begin{figure}[H]
\centering
\includegraphics[width=1\textwidth]{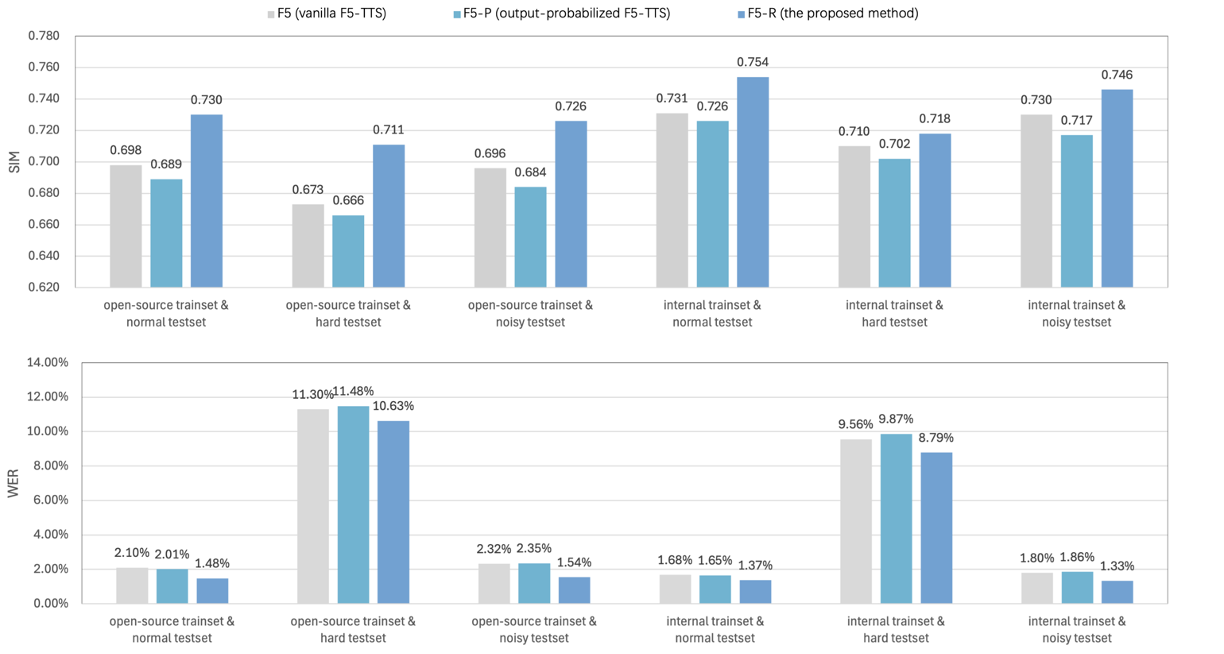}

\caption{We conducted zero-shot voice cloning experiments comparing three distinct models across different datasets. The evaluation was performed from two key perspectives: speaker similarity (measured by SIM) and semantic accuracy (measured by WER). Higher SIM and lower WER indicate superior performance.}
\label{all_result}
\end{figure}

\section{Introduction}
In recent years, significant advancements in Text-to-Speech (TTS) systems have enabled the generation of high-fidelity, natural-sounding voices and zero-shot voice cloning capabilities.These developments span both autoregressive (AR) ~\cite{wang2023neural,zhang2023speak,chen2024vall,lajszczak2024base} and non-autoregressive (NAR)~\cite{le2023voicebox,eskimez2024e2,chen2024f5} model architectures. AR models typically encode audio into discrete tokens using speech codecs and then employ language model (LM) based autoregressive models to predict these tokens. However, such approaches suffer from inference latency and exposure bias. In contrast, NAR models based on denoising diffusion or flow matching, leverage parallel computation for faster inference speeds, demonstrating strong application potential.

Additionally, as exemplified by the DeepSeek series ~\cite{shao2024deepseekmath, liu2024deepseekv2, liu2024deepseekv3, guo2025deepseekr1}, reinforcement learning (RL) has triggered a trend in the research of large language model (LLM). RL methods like Direct Preference Optimization (DPO)~\cite{rafailov2023direct} and Group Relative Policy Optimization (GRPO)~\cite{shao2024deepseekmath} have proven effective in aligning LLM outputs with human preferences, enhancing the safety and utility of generated text through feedback optimization. In the field of image generation, RL methods like denoising diffusion policy optimization (DDPO)~\cite{black2023training} have also been successfully applied. This paradigm has now extended to AR TTS systems:  Seed-TTS~\cite{anastassiou2024seed} achieved RL integration using speaker similarity (SIM) and Word Error Rate (WER) as rewards with Proximal Policy Optimization (PPO)~\cite{schulman2017proximal}, REINFORCE~\cite{ahmadian2024back} and DPO. In some other works of AR architecture, DPO and its variants are also explored~\cite{tian2024preference,gao2025emo,hussain2025koel,adler2024nemotron}.
However, integrating RL into the NAR architectures remains challenging due to fundamental structural divergences from LLMs. Current research shows no successful cases of RL integration in NAR based TTS systems, indicating that this challenge still awaits viable research solutions.


In this paper, we introduce F5R-TTS which is a novel TTS system adapting GRPO to flow-matching models through two key innovations. 
First, we reformulate the deterministic outputs of a flow-matching based model into probabilistic sequences, where F5-TTS~\cite{chen2024f5} is adopted as the backbone for our modifications.
This reformulation enables seamless integration of RL algorithms in the subsequent phase. Second, a GRPO-driven enhancement stage is designed with WER and SIM as the reward metrics, both of which are highly correlated with human perception. Experimental results demonstrate the system's effectiveness, showing significant improvements in both speech intelligibility (a 29.5\% relative reduction in WER) and speaker consistency (a 4.6\% relative increase in SIM score) compared to conventional NAR TTS baselines.

The key contributions of this work are as follows.
\begin{itemize}
    \item We propose a method to transform the outputs of flow-matching based TTS models into probabilistic representations, which enables convenient application of various reinforcement learning algorithms to flow-matching models.
    \item We successfully apply the GRPO method to NAR-TTS models, using WER and SIM as reward signals.
    \item We have implemented the F5R-TTS model in zero-shot voice cloning application scenarios, demonstrating its effectiveness.

\end{itemize}

The remainder of this paper is organized as follows: Section 2 describes the proposed method. Section 3 then presents the experiment setup and evaluation results. Finally, Section 4 concludes the paper.

\section{Proposed Method}
\label{Proposed}
The proposed method divides the training process into two phases.
We first pretrain the model with flow matching loss and subsequently improve the model with GRPO.  In this section, we will provide a detailed explanation of how the GRPO strategy is leveraged to improve the flow-matching based model.

\subsection{Preliminaries}

Our model is designed mainly following F5-TTS~\cite{chen2024f5}, which is a novel flow-matching based TTS model with zero-shot ability. The model is trained on the text-guided speech-infilling task. According to the concept of flow matching, the goal is to predict $x_{1}-x_{0}$ with $(1-t)x_{0}+tx_{1}$ as input where $x_{1} \sim $ data distribution $q(x)$ and $x_{0} \sim N(0,1)$. The vanilla objective function is defined as
\begin{equation}
L_{CFM}(\theta)=\mathbb{E}_{t,q(x_{1}),p(x_{0})}\left\Vert v_{t}((1-t)x_{0}+tx_{1})-(x_{1}-x_{0})\right\Vert_{2}
\end{equation}
where $\theta$ parameterizes the neural network $v_{t}$.

And we aim to further enhance the performance of the model using GRPO, a simplified variant of PPO that eliminates the value model and computes rewards via rule-based or model-based methods. The penalty term KL divergence $\mathbb{D}_{KL}$ between $\pi_{\theta}$ and $\pi_{ref}$ is estimated as shown in formula.~\ref{kl}.

\begin{equation}
\label{kl}
\mathbb{D}_{KL}\left[\pi_{\theta}||\pi_{ref}\right]=\frac{\pi_{ref}(o_{i,t}|q,o_{i,<t})}{\pi_{\theta}(o_{i,t}|q,o_{i,<t})}-\log\frac{\pi_{ref}(o_{i,t}|q,o_{i,<t})}{\pi_{\theta}(o_{i,t}|q,o_{i,<t})}-1,
\end{equation}

For each question $q$, it calculates the advantage based on the relative rewards of the outputs $o$ within each group and then optimizes the policy model $\pi_{\theta}$ by maximizing the following objective.

\begin{scriptsize}
\begin{equation}
\begin{split}
\mathcal{J}&_{GRPO}(\theta) =\mathbb{E}_{q\sim P(Q),\{o_{i}\}_{i=1}^{G}\sim\pi_{\theta_{old}}(O|q)} \frac{1}{G}\sum_{i=1}^{G}\frac{1}{|o_{i}|}\\
& \sum_{t=1}^{|o_{i}|}\left\{\min\left[\frac{\pi_{\theta}(o_{i,t}|q,o_{i,<t})}{\pi_{\theta_{old}}(o_{i,t}|q,o_{i,<t})}\hat{A}_{i,t},\mathrm{clip}\left(\frac{\pi_{\theta}(o_{i,t}|q,o_{i,<t})}{\pi_{\theta_{old}}(o_{i,t}|q,o_{i,<t})},1-\varepsilon,1+\varepsilon\right)\hat{A}_{i,t}\right]-\beta\mathbb{D}_{KL}\left[\pi_{\theta}||\pi_{ref}\right]\right\}
\end{split}
\end{equation}
\end{scriptsize}

where $\varepsilon$ and $\beta$ are hyper-parameters, advantage $\hat{A}_{i,t}=\widetilde{r_i}=\frac{r_i-\mathrm{mean}(\mathbf{r})}{\mathrm{std}(\mathbf{r})}$.

\subsection{Output Probabilization and Pretraining}

\begin{figure}
\centering
\includegraphics[width=0.4\textwidth]{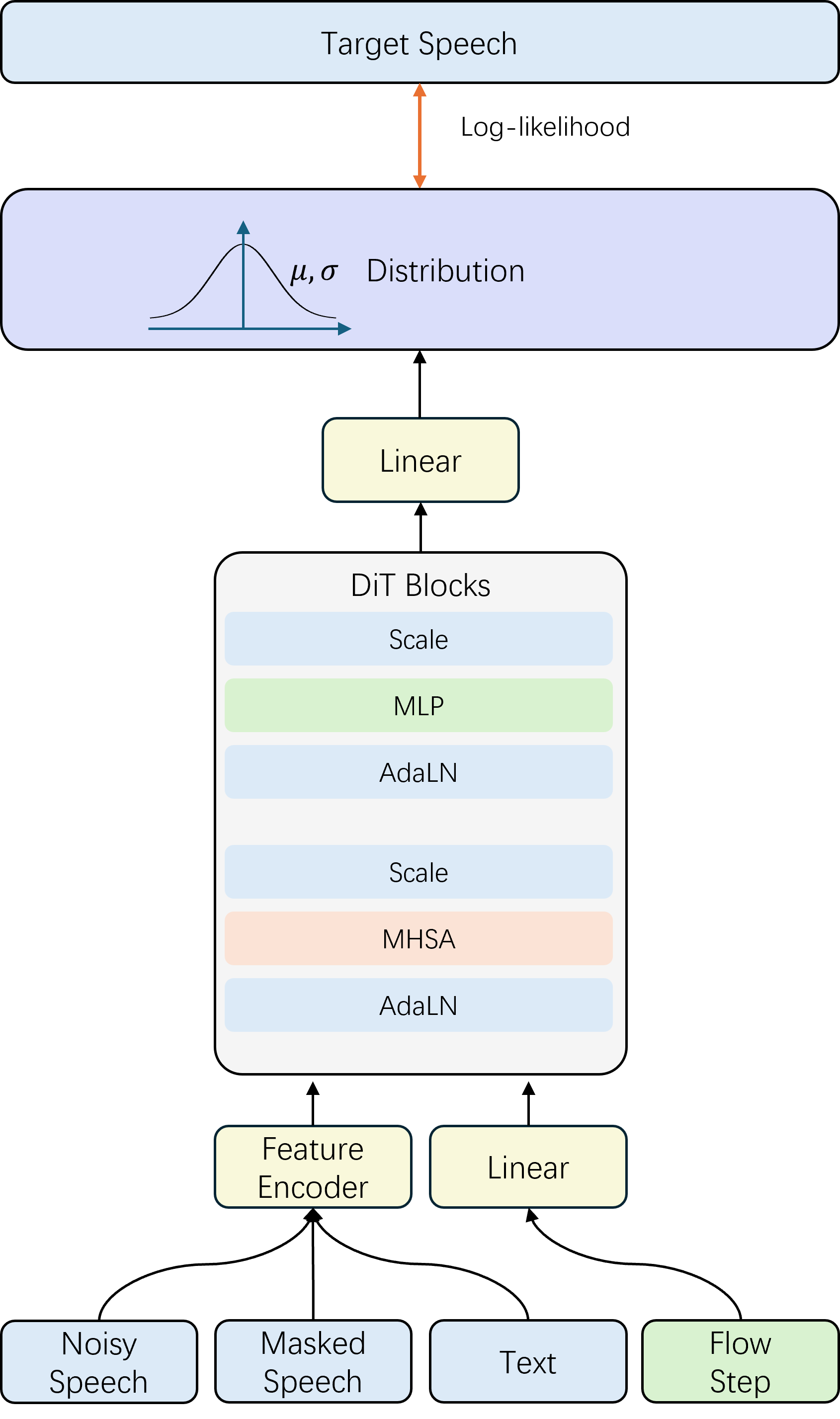} 
\caption{The backbone of F5R-TTS, which is derived from flow-matching based TTS model. 
The most significant difference in our model is the modification of the final linear layer to accurately predict probability distributions for each flow step.
}
\label{arch}
\end{figure}

We reformulate the model's output into probabilistic terms to enhance compatibility with GRPO, enabling prediction of the distribution probability for $x_{1}-x_{0}$. Fig.~\ref{arch} shows the overall structure of the model. In the first phase, we retain the flow matching objective function. The flow matching objective is to match a probability path from a standard normal distribution to a distribution approximating the data distribution.

The proposed model is trained on the infilling task as well. During training, the model takes the flow step $t$, the noisy acoustic feature $(1-t)x_{0} + tx_{1}$, the masked acoustic feature $(1-m) \odot x_{1}$, and the transcript text of complete speech $T_{gt}$ as inputs. We use extracted mel-spectrogram feature as acoustic feature for training and pad the text feature to the same length as the acoustic feature.

The proposed model does not directly predict the exact value of $m \odot (x_{1}-x_{0})$.
We let the model predict the mean $\mu(x)$ and variance $\sigma(x)$ of the Gaussian distribution in the last layer, and the parameters $\theta$ are optimized to maximize the following log-likelihood of $x_{1}-x_{0}$.
\begin{equation}
\mathbb{J}_{CFM}(\theta)=\mathbb{E}_{t,q(x_{1}),p(x_{0})} \mathrm{log}p_{\theta}((x_{1}-x_{0})|((1-t)x_{0}+tx_{1}))
\label{j_cfm}
\end{equation}

After simplifying the formula.~\ref{j_cfm}, we can get the following modified objective function
\begin{equation}
L_{CFM}(\theta)=\mathbb{E}_{t,q(x_{1}),p(x_{0})}\left[ \frac {(\mu((1-t)x_{0}+tx_{1})-(x_{1}-x_{0}))^{2}}{2\sigma((1-t)x_{0}+tx_{1})^{2}}+\mathrm{log}\sigma((1-t)x_{0}+tx_{1})\right]
\label{l_cfm}
\end{equation}

During the pretraining phase, we use formula.~\ref{l_cfm} to optimize the model.

\subsection{Model Enhancement with GRPO}

\begin{figure}
\centering
\includegraphics[width=1.0\textwidth]{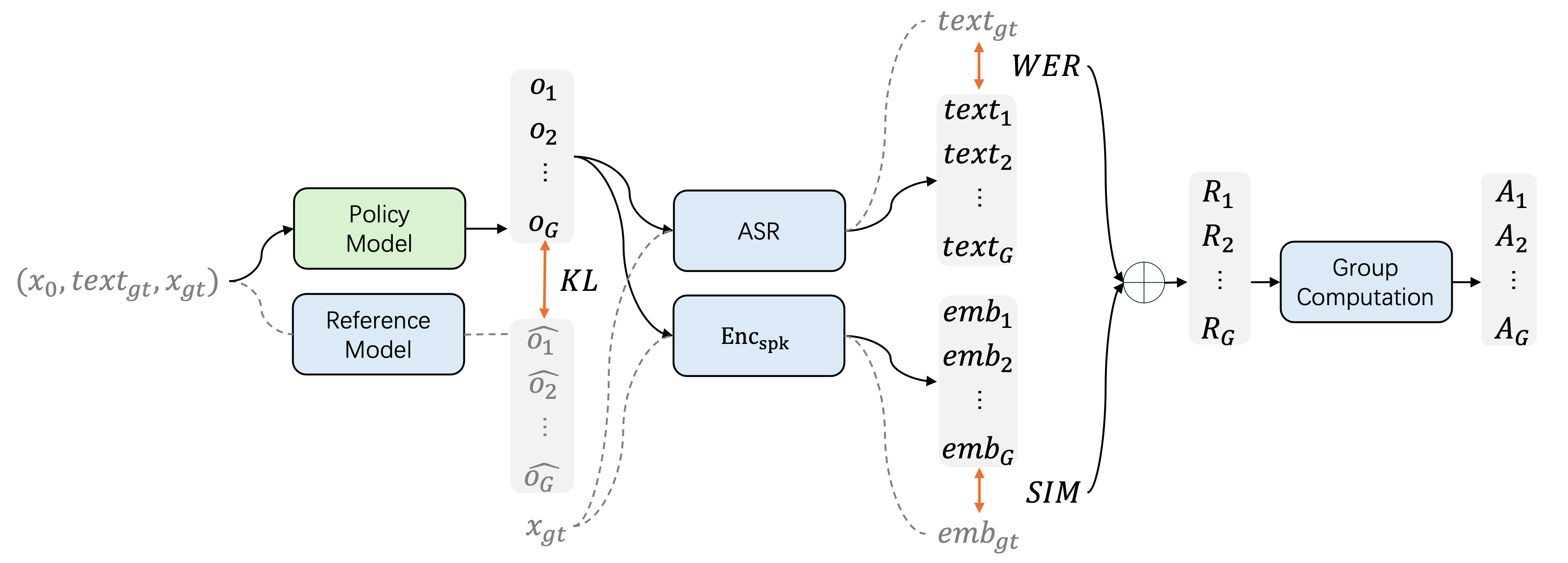} 
\caption{The pipeline of the GRPO phase. We employ an ASR model and a speaker encoder to derive rewards, which is subsequently used to optimize the policy model. KL divergence is incorporated as the penalty term to enhance training stability during GRPO phase.
}
\label{grpo}
\end{figure}

We continue to use GRPO to improve the performance of the model after the pretraining phase. The pipeline of the GRPO phase is shown in Fig.~\ref{grpo}

In the second phase, we further train the pretrained model as policy model $\pi_{\theta}$ while initializing the reference model $\pi_{ref}$ with the pretrained parameters. The reference model remains frozen throughout GRPO phase. The forward operation of our TTS model during GRPO training is different from the pretraining phase. There is a sampling operation that is similar to inference. Policy model $\pi_{\theta}$ takes $x_{0}\sim N(0,1)$ as input, and then the output probability is calculated for each flow step. The sampling result of policy model $o$ is used to calculate rewards and the KL loss compared with the reference model result $x_{ref}$.

In terms of reward metrics, we select WER and SIM as the primary criteria to improve semantic consistency and speaker similarity, as these represent the two most critical aspects in voice cloning tasks. We employ an ASR model to transcribe the synthesized speech, obtain the transcribed text $T_{pol}$, and then compare the transcriptions with the ground truth text $T_{gt}$ of the ground truth speech to calculate WER. Additionally, we utilize a speaker encoder $Enc_{spk}$ to extract synthesized speaker embedding $emb_{pol}$ and ground truth speaker embedding $emb_{gt}$ from the generated speech $o$ and ground truth speech sample $x_{gt}$ respectively. Speaker similarity is evaluated by computing cosine similarity between these embeddings.

Therefore, GRPO reward is divided into semantic-related reward and speaker-related reward, which are defined as follows.

\begin{equation}
Reward_{W}=\mathbb{E} \left[1-\mathit{WER}(T_{ft},T_{pol})\right]
\label{reward_wer}
\end{equation}

\begin{equation}
Reward_{S}=\mathbb{E} \left[\mathit{SIM}(emb_{gt},emb_{pol})\right]
\label{reward_sim}
\end{equation}

The entire reward is defined as

\begin{equation}
Reward=\lambda_{W}Reward_{W} + \lambda_{S}Reward_{S}
\label{reward_grpo}
\end{equation}
where $\lambda_{W}$ and $\lambda_{S}$ are the weight items for respective reward.

After calculating the reward, we can get the advantage through group relative advantage estimation~\cite{shao2024deepseekmath}.

\begin{equation}
A_{i}=\frac{Reward_{i} - mean(Reward)}{std(Reward)}
\label{adv}
\end{equation}

In order to maintain the stability of the model output, GRPO also needs to use the reference model $\pi_{ref}$ to provide constraints. Finally, we define formula.~\ref{l_grpo} as the objective function in the second phase.

\begin{equation}
max\mathbb{E}_{x_{0}\sim N(0,1),o_{i}\sim \pi_{\theta}(o|x_{0},text_{gt})}\left[ \frac {1}{G}\sum_{i=1}^{G}\pi_{\theta}(o_{i}|x_{0},text_{gt})A_{i}-\beta\mathbb{D}_{KL}(\pi_{\theta} \Vert \pi_{ref})\right]
\label{l_grpo}
\end{equation}

\section{Experiments}
\label{exp}

In this section, our experiments are focused on validating the efficacy of the proposed method in enhancing performance on zero-shot voice cloning tasks.

\subsection{Dataset and Experiment Setup}

For the pretraining phase, we utilized WenetSpeech4TTS Basic~\cite{ma2024wenetspeech4tts}, a Mandarin open-source dataset consisting of 7,226 hours of multi-speaker corpus, as the training set. During the GRPO phase, we randomly selected 100 hours of speech data from the same dataset for training. In the evaluation, following the test setting of Seed-TTS, we synthesized 2,020 general samples and 400 hard samples using the reference speech from the Seed-TTS-eval test-cn set\footnote{\href{https://github.com/BytedanceSpeech/seed-tts-eval}{https://github.com/BytedanceSpeech/seed-tts-eval}}. General samples utilize plain text, whereas hard samples employ text that presents difficulties, such as tongue twisters or containing a high frequency of repetitive words and phrases. To test noise robustness, we generated 140 samples using 70 noisy utterances from the same test set.

Our model architecture was built mainly following the configuration described in the F5-TTS paper, with modifications applied only to the last output layer. During the pretraining phase, the model was trained for 1 million updates on 8 A100 40GB GPUs with a batch size of 160,000 frames. In the GRPO training phase, the model was trained for 1,100 updates on 8 A100 40GB GPUs with a batch size of 6,400 frames. For the GRPO training, we utilized SenseVoice~\cite{an2024funaudiollm} as the automatic speech recognition (ASR) system to compute the $Reward_{W}$ and WeSpeaker~\cite{wang2023wespeaker} as the speaker encoder for $Reward_{S}$ calculation.

We selected vanilla F5-TTS as the baseline for our experiments. To demonstrate the effectiveness of GRPO in improving TTS models, we compared the performance of vanilla F5-TTS, output-probabilized F5-TTS, and the proposed method. We named them F5, F5-P, and F5-R respectively. The F5 strictly preserves the original architecture and parameter settings. F5-P is also the pretrained model of the GRPO phase. We trained all models on the identical pretraining dataset.

\subsection{Evaluation}

\subsubsection{Visualization Analysis}

We try to visualize the differences in the performance of different models in zero-shot voice cloning tasks.

\begin{figure}
\centering
\begin{minipage}{0.3\textwidth}
\centering
\includegraphics[width=1\textwidth]{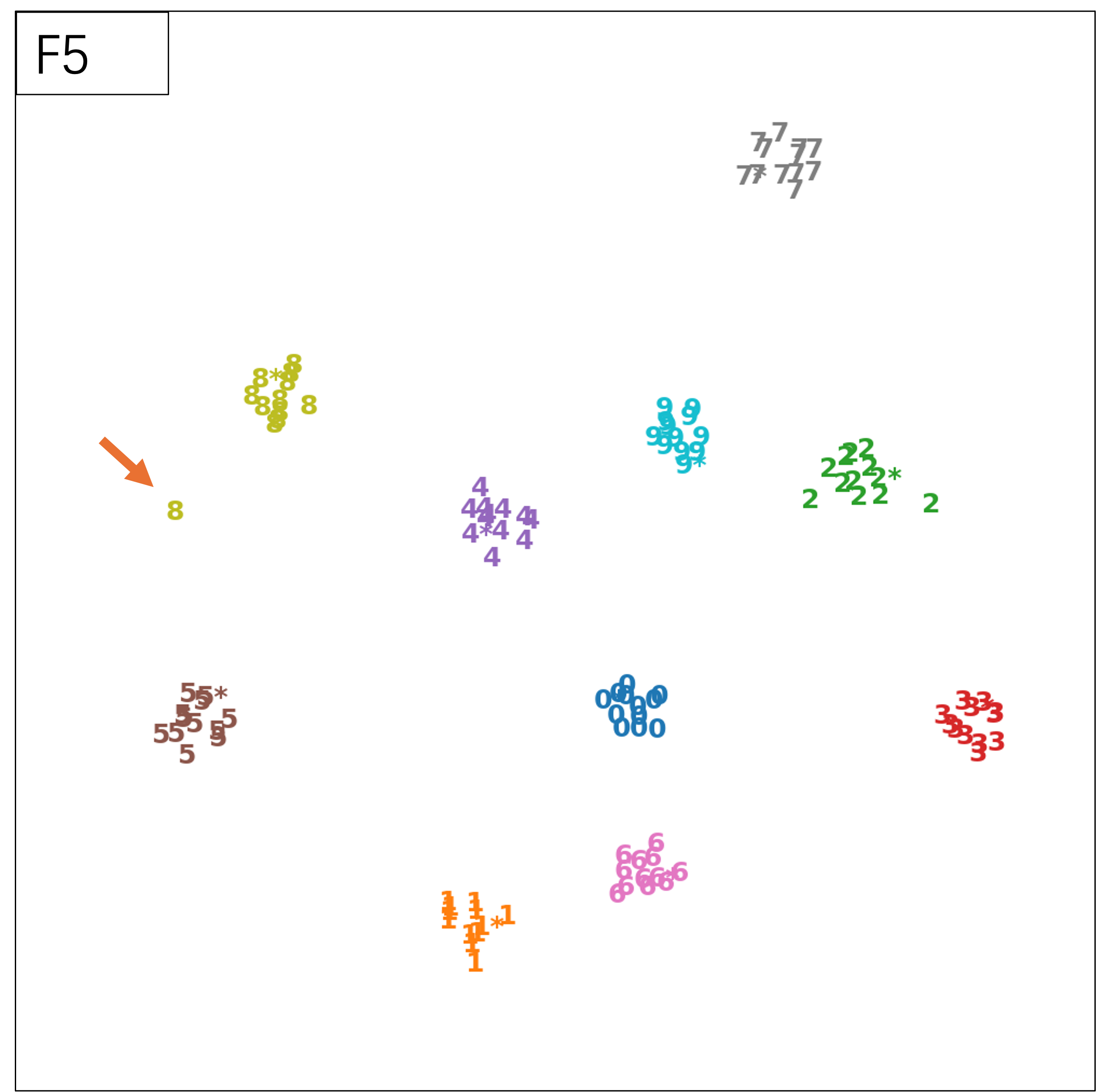}
\end{minipage}%
\begin{minipage}{0.3\textwidth}
\centering
\includegraphics[width=1\textwidth]{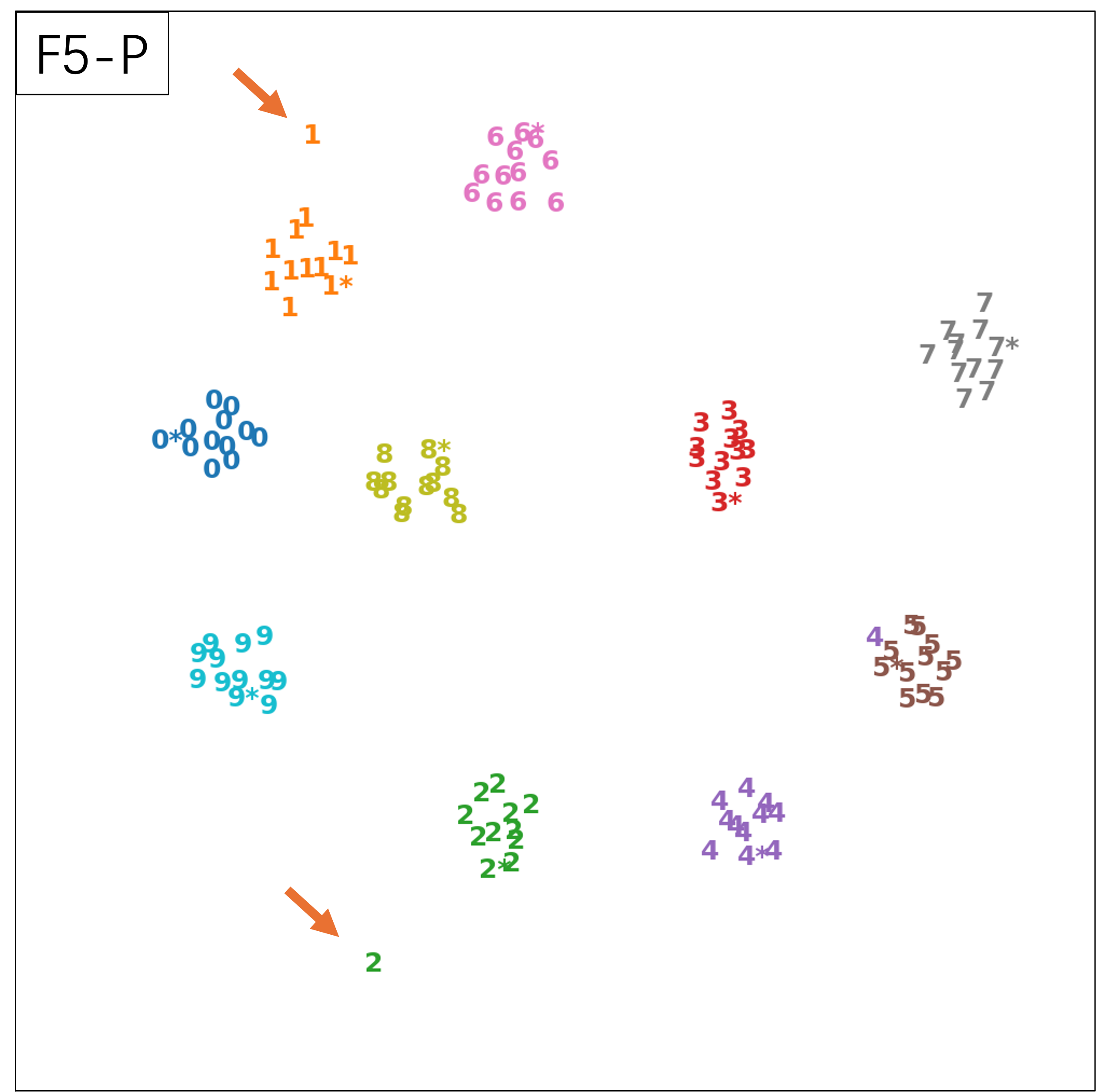}
\end{minipage}%
\begin{minipage}{0.3\textwidth}
\centering
\includegraphics[width=1\textwidth]{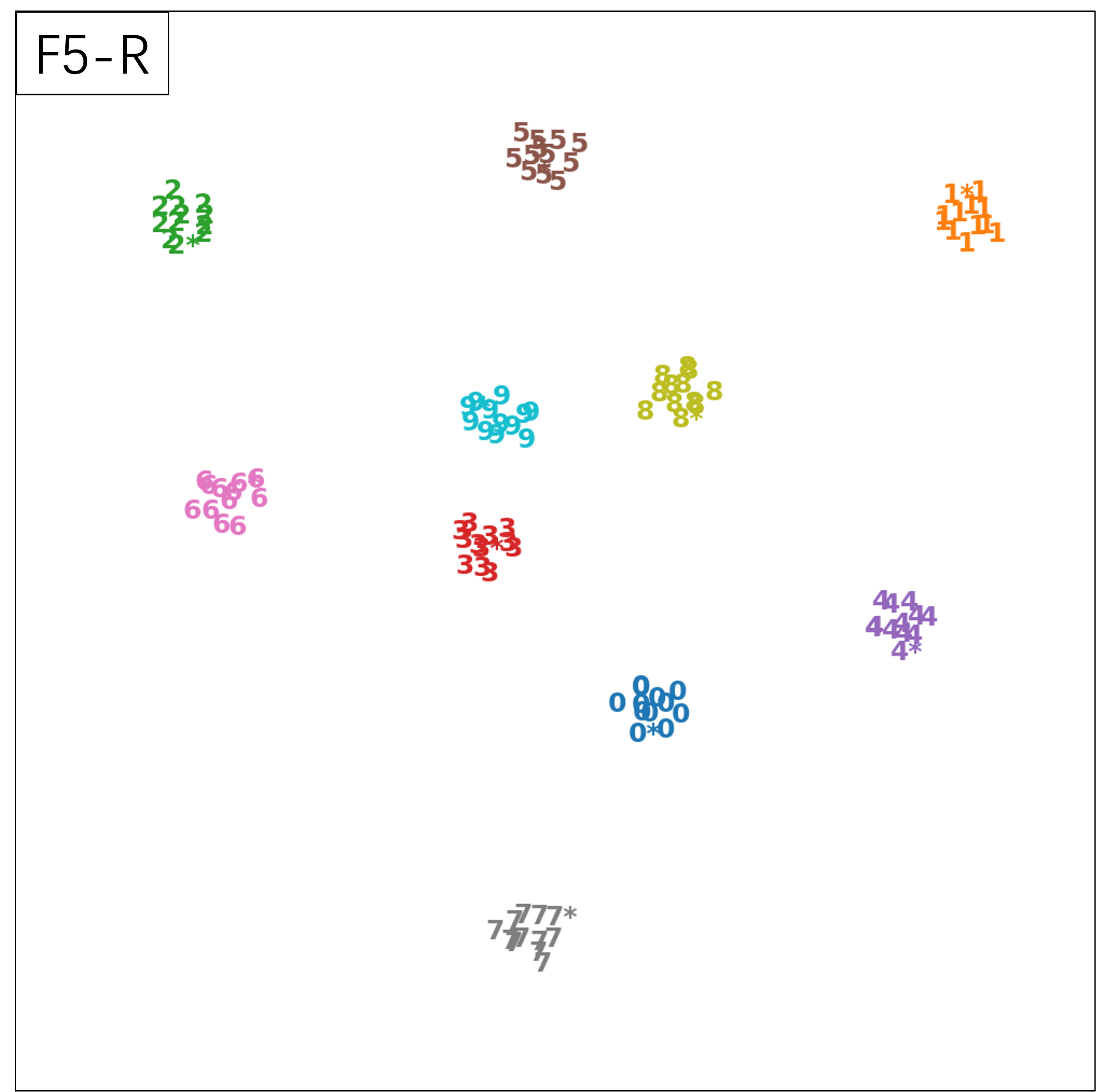}
\end{minipage}%

\caption{The visualization of speaker similarity by t-SNE. From left to right, three columns correspond to F5, F5-P, and F5-R, respectively. Each small number in the graph is an utterance sample. Different numbers or colors correspond to different target speakers. Numbers with an asterisk mean reference utterances of the target speaker whom the number stands for. Numbers without an asterisk refer to synthesized utterances. And some badcases are marked out with red arrows.}
\label{t_sne}
\end{figure}

\begin{figure}
\centering
\begin{minipage}{0.4\textwidth}
\centering
\includegraphics[width=1\textwidth]{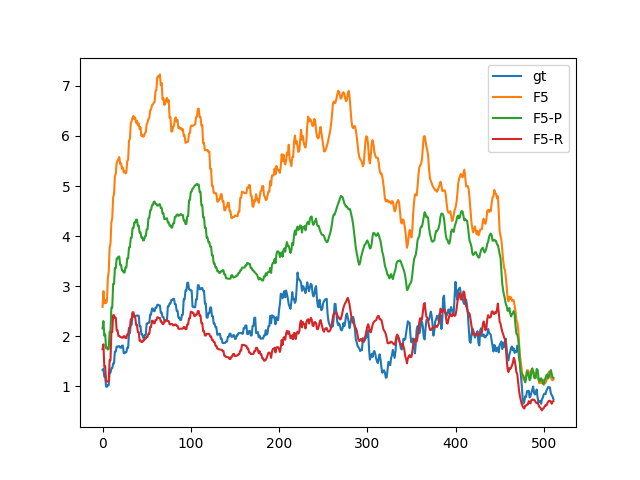}
\end{minipage}%
\begin{minipage}{0.4\textwidth}
\centering
\includegraphics[width=1\textwidth]{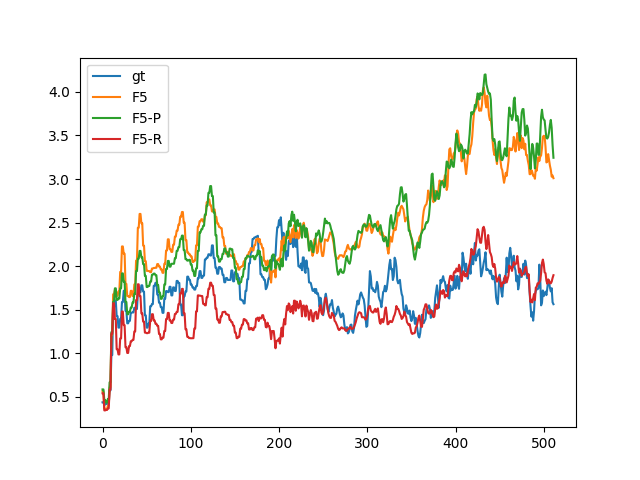}
\end{minipage}%

\begin{minipage}{0.4\textwidth}
\centering
\includegraphics[width=1\textwidth]{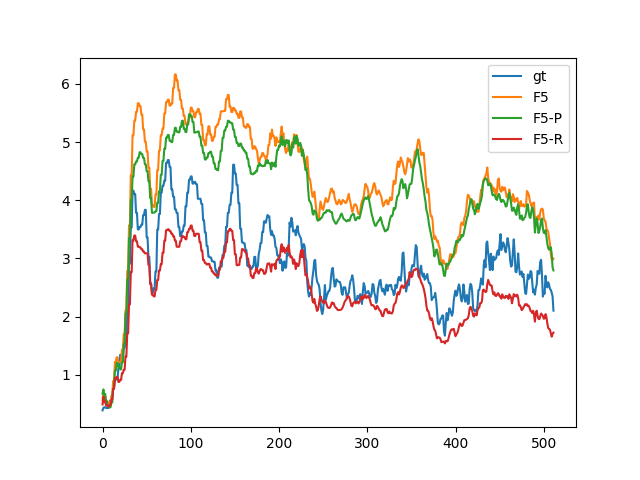}
\end{minipage}%
\begin{minipage}{0.4\textwidth}
\centering
\includegraphics[width=1\textwidth]{figures/gv-10002309-00000029.png}
\end{minipage}%

\caption{The global variance of ground truth speaker utterance and synthesized utterances from different models. In each subgraph, the horizontal axis represents the mel bins number and the vertical axis represents the variance. And there are 4 GV curves in each subgraph corresponding to different sources. The corresponding relationship of the curves is shown in the legend where gt means ground truth.}
\label{gv}
\end{figure}

We first used t-SNE~\cite{2008Visualizing} to visualize the speaker similarity in the 2D space. T-SNE can cluster data, such as speaker embeddings, in an unsupervised manner. For this analysis, we randomly selected 10 unseen speakers from Seed-TTS-eval test-cn set as target speakers, and then the models respectively synthesized 10 utterances per speaker. WeSpeaker was utilized to get speaker embeddings of utterances, which were then visualized via t-SNE. We can intuitively see the similarity between the synthesized results and the ground truth samples. We can also observe distribution differences across target speakers. As shown in Fig.\ref{t_sne}, the results of F5-R were well clustered according to the target speaker. Meanwhile, the subgraph of F5 and F5-P shows that the synthesis results corresponding to some target speakers were not completely clustered together. This means that the synthesis results of F5-R have better speaker similarity.

Secondly, we used the global variance (GV)~\cite{2005Spectral}. GV is a method to visualize the spectral variance distribution of utterances. We respectively generated 20 utterances for 4 unseen speakers (2 females and 2 males) of the Seed-TTS-eval test-cn set. We then calculated GV for both reference and synthesized utterances. A closer match between the synthesized and reference curves indicates better performance. As shown in Fig.~\ref{gv}, the red curve for F5-R shows better alignment with the blue curve for reference utterances than others, which also shows that the synthesized results of F5-R are more similar to reference utterances.

\subsubsection{Metrics Analysis}

\begin{table}
  \caption{Objective metrics comparison for zero-shot voice cloning among different methods trained with WenetSpeech4TTS Basic.}
  \label{obj}
  \centering
  \begin{tabular}{c|c|c|c}
    \toprule
    Method & test set & WER & SIM \\
    \midrule
    F5 & ZH  & 2.10\% & 0.698     \\
    \cline{2-4}
     & hard  & 11.30\% & 0.673     \\
    \cline{2-4}
     & noisy  & 2.32\% & 0.696     \\
    \midrule
    F5-P & ZH  & 2.01\% & 0.689     \\
    \cline{2-4}
     & hard  & 11.48\% & 0.666     \\
    \cline{2-4}
     & noisy  & 2.35\% & 0.684     \\
    \midrule
    F5-R & ZH  & 1.48\% & 0.730     \\
    \cline{2-4}
     & hard  & 10.63\% & 0.711     \\
    \cline{2-4}
     & noisy  & 1.54\% & 0.726     \\
    \bottomrule
  \end{tabular}
\end{table}

To evaluate model performance, we adopted WER and SIM as objective metrics, based on seed-tts-eval test-cn. For metric computation, we leveraged the official evaluation toolkit provided with seed-tts-eval. For WER, we employed Paraformer-zh~\cite{gao2023funasr} to transcribe. For SIM, we utilized a WavLM-large-based speaker verification model~\cite{chen2022large} to extract speaker embedding. These metrics respectively quantify semantic accuracy (lower WER preferred) and speaker similarity (higher SIM desired). Table.~\ref{obj} presents the comparative evaluation results across the three models.

In terms of SIM, it can be observed that F5 and F5-P exhibit comparable performance across two distinct test sets, with F5 marginally outperforming the latter. Our proposed model achieves superior performance on both sets, establishing itself as the top performer. Notably, on the general test set, our model outperforms others with a minimum advantage of 0.03 SIM points. Compared to F5, F5-R achieved a 4.6\% and 5.6\% relative increase on the general test set and the hard test set respectively. This indicates that GRPO contributes positively to improving speaker similarity.

In terms of WER, F5 and F5-P remain closely aligned across the two distinct test sets. Our model, however, achieves significantly better results on both sets. On the general test set, it attains a 29.5\% relative WER reduction compared to the baseline, with a further 5.9\% reduction observed on the hard test set. These results conclusively confirm the effectiveness of GRPO in enhancing semantic accuracy.

For the noise robustness test, we used noisy utterances as reference audio. Noise exerted negligible influence on SIM across all models, while inducing an increase in the WER of F5 and F5-P. Meanwhile, F5-R demonstrated significantly improved noise robustness. And we found that F5-R also maintained superior rhythm performance under noise conditions. Audio samples of noise robustness test are available at \url{https://frontierlabs.github.io/F5R}.

\begin{table}
  \caption{Objective metrics comparison for zero-shot voice cloning among different methods trained with our internal dataset.}
  \label{obj2}
  \centering
  \begin{tabular}{c|c|c|c}
    \toprule
    Method & test set & WER & SIM \\
    \midrule
    F5 & ZH  & 1.68\% & 0.731     \\
    \cline{2-4}
     & hard  & 9.56\% & 0.710     \\
    \cline{2-4}
     & noisy  & 1.80\% & 0.730     \\
    \midrule
    F5-P & ZH  & 1.65\% & 0.726     \\
    \cline{2-4}
     & hard  & 9.87\% & 0.702     \\
    \cline{2-4}
     & noisy  & 1.86\% & 0.717     \\
    \midrule
    F5-R & ZH  & 1.37\% & 0.754     \\
    \cline{2-4}
     & hard  & 8.79\% & 0.718     \\
    \cline{2-4}
     & noisy  & 1.33\% & 0.746     \\
    \bottomrule
  \end{tabular}
\end{table}

To demonstrate the generalizability of the proposed method, we conducted parallel experiments on additional datasets. For the pretraining phase, we employed an internal Mandarin dataset comprising 10,000 hours of corpus, primarily sourced from radio broadcasts and audiobooks. We conducted a preliminary quality-based filtering for the dataset. Furthermore, a 100-hour subset was randomly selected for GRPO training. The objective metrics comparison with the internal dataset is shown in Table.~\ref{obj2}. Compared to F5, F5-R achieved an 18.4\% relative reduction in and a 3.1\% relative increase in SIM on the general test set. On the hard test set, F5-R achieved a WER relative reduction of 8.1\% and a SIM relative increase of 1.1\%. On the noisy test set, F5-R achieved a WER relative reduction of 26.1\% and a SIM relative increase of 2.2\%. The overall results align with those obtained on WenetSpeech4TTS Basic, showing that GRPO consistently enhances model performance across diverse datasets.

Overall, the performance of F5 and F5-P remains largely comparable. As anticipated, the GRPO with WER and SIM as rewards enables the model to achieve gains in both semantic accuracy and speaker similarity. Guided by the speaker-related component of the reward, the model demonstrates enhanced capability to clone target speaker characteristics through in-context learning. On the hard test set, the proposed model exhibits more pronounced relative advantages in WER performance. We hypothesize that this improvement stems from the WER-related reward component, which effectively strengthens the model's semantic preservation ability. However, all three models show performance degradation on the hard test set, suggesting that increased text complexity generally reduces model stability. This observation may serve as a key focus for future optimization efforts.

\section{Conclusion}
In this paper, we propose F5R-TTS, which introduces the GRPO method into a flow matching-based NAR TTS system. By transforming the outputs of flow-matching based model into probabilistic representations, GRPO can be integrated into the training pipeline.  Experimental results demonstrate that the proposed method achieves higher SIM and lower WER compared to baseline systems, indicating that GRPO with appropriate reward functions positively contributes to both semantic accuracy and speaker similarity. 

Our next step involves investing in research across the following directions.
\begin{itemize}
    \item \textbf{RL approach investigation}: We plan to explore the integration of additional reinforcement learning approaches (e.g., PPO, DDPO) into NAR TTS systems.
    \item \textbf{Reward function optimization}: In order to further enhance the model's stability in challenging scenarios, we will continue to investigate optimized reward function designs.
    \item \textbf{Data exploration}: To better understand the model's performance with larger datasets, we will utilize more training data for further experiments.
\end{itemize}

\bibliography{F5R-TTS}

\begin{thebibliography}{10}

\bibitem{wang2023neural}
Chengyi Wang, Sanyuan Chen, Yu~Wu, Ziqiang Zhang, Long Zhou, Shujie Liu, Zhuo Chen, Yanqing Liu, Huaming Wang, Jinyu Li, et~al.
\newblock Neural codec language models are zero-shot text to speech synthesizers.
\newblock {\em arXiv preprint arXiv:2301.02111}, 2023.

\bibitem{zhang2023speak}
Ziqiang Zhang, Long Zhou, Chengyi Wang, Sanyuan Chen, Yu~Wu, Shujie Liu, Zhuo Chen, Yanqing Liu, Huaming Wang, Jinyu Li, et~al.
\newblock Speak foreign languages with your own voice: Cross-lingual neural codec language modeling.
\newblock {\em arXiv preprint arXiv:2303.03926}, 2023.

\bibitem{chen2024vall}
Sanyuan Chen, Shujie Liu, Long Zhou, Yanqing Liu, Xu~Tan, Jinyu Li, Sheng Zhao, Yao Qian, and Furu Wei.
\newblock Vall-e 2: Neural codec language models are human parity zero-shot text to speech synthesizers.
\newblock {\em arXiv preprint arXiv:2406.05370}, 2024.

\bibitem{lajszczak2024base}
Mateusz {\L}ajszczak, Guillermo C{\'a}mbara, Yang Li, Fatih Beyhan, Arent Van~Korlaar, Fan Yang, Arnaud Joly, {\'A}lvaro Mart{\'\i}n-Cortinas, Ammar Abbas, Adam Michalski, et~al.
\newblock Base tts: Lessons from building a billion-parameter text-to-speech model on 100k hours of data.
\newblock {\em arXiv preprint arXiv:2402.08093}, 2024.

\bibitem{le2023voicebox}
Matthew Le, Apoorv Vyas, Bowen Shi, Brian Karrer, Leda Sari, Rashel Moritz, Mary Williamson, Vimal Manohar, Yossi Adi, Jay Mahadeokar, et~al.
\newblock Voicebox: Text-guided multilingual universal speech generation at scale.
\newblock {\em Advances in neural information processing systems}, 36:14005--14034, 2023.

\bibitem{eskimez2024e2}
Sefik~Emre Eskimez, Xiaofei Wang, Manthan Thakker, Canrun Li, Chung-Hsien Tsai, Zhen Xiao, Hemin Yang, Zirun Zhu, Min Tang, Xu~Tan, et~al.
\newblock E2 tts: Embarrassingly easy fully non-autoregressive zero-shot tts.
\newblock In {\em 2024 IEEE Spoken Language Technology Workshop (SLT)}, pages 682--689. IEEE, 2024.

\bibitem{chen2024f5}
Yushen Chen, Zhikang Niu, Ziyang Ma, Keqi Deng, Chunhui Wang, Jian Zhao, Kai Yu, and Xie Chen.
\newblock F5-tts: A fairytaler that fakes fluent and faithful speech with flow matching.
\newblock {\em arXiv preprint arXiv:2410.06885}, 2024.

\bibitem{shao2024deepseekmath}
Zhihong Shao, Peiyi Wang, Qihao Zhu, Runxin Xu, Junxiao Song, Xiao Bi, Haowei Zhang, Mingchuan Zhang, YK~Li, Y~Wu, et~al.
\newblock Deepseekmath: Pushing the limits of mathematical reasoning in open language models.
\newblock {\em arXiv preprint arXiv:2402.03300}, 2024.

\bibitem{liu2024deepseekv2}
Aixin Liu, Bei Feng, Bin Wang, Bingxuan Wang, Bo~Liu, Chenggang Zhao, Chengqi Dengr, Chong Ruan, Damai Dai, Daya Guo, et~al.
\newblock Deepseek-v2: A strong, economical, and efficient mixture-of-experts language model.
\newblock {\em arXiv preprint arXiv:2405.04434}, 2024.

\bibitem{liu2024deepseekv3}
Aixin Liu, Bei Feng, Bing Xue, Bingxuan Wang, Bochao Wu, Chengda Lu, Chenggang Zhao, Chengqi Deng, Chenyu Zhang, Chong Ruan, et~al.
\newblock Deepseek-v3 technical report.
\newblock {\em arXiv preprint arXiv:2412.19437}, 2024.

\bibitem{guo2025deepseekr1}
Daya Guo, Dejian Yang, Haowei Zhang, Junxiao Song, Ruoyu Zhang, Runxin Xu, Qihao Zhu, Shirong Ma, Peiyi Wang, Xiao Bi, et~al.
\newblock Deepseek-r1: Incentivizing reasoning capability in llms via reinforcement learning.
\newblock {\em arXiv preprint arXiv:2501.12948}, 2025.

\bibitem{rafailov2023direct}
Rafael Rafailov, Archit Sharma, Eric Mitchell, Christopher~D Manning, Stefano Ermon, and Chelsea Finn.
\newblock Direct preference optimization: Your language model is secretly a reward model.
\newblock {\em Advances in Neural Information Processing Systems}, 36:53728--53741, 2023.

\bibitem{black2023training}
Kevin Black, Michael Janner, Yilun Du, Ilya Kostrikov, and Sergey Levine.
\newblock Training diffusion models with reinforcement learning.
\newblock In {\em ICML 2023 Workshop on Structured Probabilistic Inference $\{$$\backslash$\&$\}$ Generative Modeling}, 2023.

\bibitem{anastassiou2024seed}
Philip Anastassiou, Jiawei Chen, Jitong Chen, Yuanzhe Chen, Zhuo Chen, Ziyi Chen, Jian Cong, Lelai Deng, Chuang Ding, Lu~Gao, et~al.
\newblock Seed-tts: A family of high-quality versatile speech generation models.
\newblock {\em arXiv preprint arXiv:2406.02430}, 2024.

\bibitem{schulman2017proximal}
John Schulman, Filip Wolski, Prafulla Dhariwal, Alec Radford, and Oleg Klimov.
\newblock Proximal policy optimization algorithms.
\newblock {\em arXiv preprint arXiv:1707.06347}, 2017.

\bibitem{ahmadian2024back}
Arash Ahmadian, Chris Cremer, Matthias Gall{\'e}, Marzieh Fadaee, Julia Kreutzer, Olivier Pietquin, Ahmet {\"U}st{\"u}n, and Sara Hooker.
\newblock Back to basics: Revisiting reinforce-style optimization for learning from human feedback in llms.
\newblock In {\em Proceedings of the 62nd Annual Meeting of the Association for Computational Linguistics (Volume 1: Long Papers)}, pages 12248--12267, 2024.

\bibitem{tian2024preference}
Jinchuan Tian, Chunlei Zhang, Jiatong Shi, Hao Zhang, Jianwei Yu, Shinji Watanabe, and Dong Yu.
\newblock Preference alignment improves language model-based tts.
\newblock {\em arXiv preprint arXiv:2409.12403}, 2024.

\bibitem{gao2025emo}
Xiaoxue Gao, Chen Zhang, Yiming Chen, Huayun Zhang, and Nancy~F Chen.
\newblock Emo-dpo: Controllable emotional speech synthesis through direct preference optimization.
\newblock In {\em ICASSP 2025-2025 IEEE International Conference on Acoustics, Speech and Signal Processing (ICASSP)}, pages 1--5. IEEE, 2025.

\bibitem{hussain2025koel}
Shehzeen Hussain, Paarth Neekhara, Xuesong Yang, Edresson Casanova, Subhankar Ghosh, Mikyas~T Desta, Roy Fejgin, Rafael Valle, and Jason Li.
\newblock Koel-tts: Enhancing llm based speech generation with preference alignment and classifier free guidance.
\newblock {\em arXiv preprint arXiv:2502.05236}, 2025.

\bibitem{adler2024nemotron}
Bo~Adler, Niket Agarwal, Ashwath Aithal, Dong~H Anh, Pallab Bhattacharya, Annika Brundyn, Jared Casper, Bryan Catanzaro, Sharon Clay, Jonathan Cohen, et~al.
\newblock Nemotron-4 340b technical report.
\newblock {\em arXiv preprint arXiv:2406.11704}, 2024.

\bibitem{ma2024wenetspeech4tts}
Linhan Ma, Dake Guo, Kun Song, Yuepeng Jiang, Shuai Wang, Liumeng Xue, Weiming Xu, Huan Zhao, Binbin Zhang, and Lei Xie.
\newblock Wenetspeech4tts: A 12,800-hour mandarin tts corpus for large speech generation model benchmark.
\newblock In {\em Proc. Interspeech 2024}, pages 1840--1844, 2024.

\bibitem{an2024funaudiollm}
Keyu An, Qian Chen, Chong Deng, Zhihao Du, Changfeng Gao, Zhifu Gao, Yue Gu, Ting He, Hangrui Hu, Kai Hu, et~al.
\newblock Funaudiollm: Voice understanding and generation foundation models for natural interaction between humans and llms.
\newblock {\em arXiv preprint arXiv:2407.04051}, 2024.

\bibitem{wang2023wespeaker}
Hongji Wang, Chengdong Liang, Shuai Wang, Zhengyang Chen, Binbin Zhang, Xu~Xiang, Yanlei Deng, and Yanmin Qian.
\newblock Wespeaker: A research and production oriented speaker embedding learning toolkit.
\newblock In {\em ICASSP 2023-2023 IEEE International Conference on Acoustics, Speech and Signal Processing (ICASSP)}, pages 1--5. IEEE, 2023.

\bibitem{2008Visualizing}
Laurens Van~Der Maaten and Geoffrey Hinton.
\newblock Visualizing data using t-{SNE}.
\newblock {\em Journal of Machine Learning Research}, 9(2605):2579--2605, 2008.

\bibitem{2005Spectral}
Tomoki Toda, Alan~W Black, and Keiichi Tokuda.
\newblock Spectral conversion based on maximum likelihood estimation considering global variance of converted parameter.
\newblock In {\em ICASSP 2005-2005 IEEE International Conference on Acoustics, Speech, and Signal Processing (ICASSP).}, volume~1, pages I--9. IEEE, 2005.

\bibitem{gao2023funasr}
Zhifu Gao, Zerui Li, Jiaming Wang, Haoneng Luo, Xian Shi, Mengzhe Chen, Yabin Li, Lingyun Zuo, Zhihao Du, and Shiliang Zhang.
\newblock Funasr: A fundamental end-to-end speech recognition toolkit.
\newblock In {\em Proc. Interspeech 2023}, pages 1593--1597, 2023.

\bibitem{chen2022large}
Zhengyang Chen, Sanyuan Chen, Yu~Wu, Yao Qian, Chengyi Wang, Shujie Liu, Yanmin Qian, and Michael Zeng.
\newblock Large-scale self-supervised speech representation learning for automatic speaker verification.
\newblock In {\em ICASSP 2022-2022 IEEE International Conference on Acoustics, Speech and Signal Processing (ICASSP)}, pages 6147--6151. IEEE, 2022.

\end{thebibliography}

\end{document}